\documentclass[osajnl,preprint,showpacs,superscriptaddress,12pt]{revtex4-1} %% use 12pt for preprint option
\usepackage{amsmath,amssymb,graphicx}
\usepackage{epstopdf}

\begin{document}

\title{Confocal Supercritical Angle Microscopy for cell membrane imaging.}

\author{Siddharth Sivankutty}
\affiliation{Institut des Sciences Mol\'{e}culaires d'Orsay, Universit\'{e} Paris-Sud, CNRS UMR 8214, F-91405 Orsay cedex, France.}
\affiliation{Centre de photonique Biom\'{e}dicale (CLUPS), Universit\'{e} Paris-Sud, CNRS FR 2764, F-91405 Orsay cedex, France.}

\author{Thomas Barroca}
\affiliation{Institut Langevin, CNRS UMR 7587, 1 rue Jussieu, 75 238 Paris Cedex 05, France.}
\author {C\'{e}line Mayet}
\affiliation{Centre de photonique Biom\'{e}dicale (CLUPS), Universit\'{e} Paris-Sud, CNRS FR 2764, F-91405 Orsay cedex, France.}

\author{Guillaume Dupuis}
\affiliation{Centre de photonique Biom\'{e}dicale (CLUPS), Universit\'{e} Paris-Sud, CNRS FR 2764, F-91405 Orsay cedex, France.}

\author{Emmanuel Fort} \email{emmanuel.fort@espci.fr}
\affiliation{Institut Langevin, CNRS UMR 7587, 1 rue Jussieu, 75 238 Paris Cedex 05, France.}

\author{Sandrine L\'{e}v\^{e}que-Fort} \email{sandrine.leveque-fort@u-psud.fr}
\affiliation{Institut des Sciences Mol\'{e}culaires d'Orsay, Universit\'{e} Paris-Sud, CNRS UMR 8214, F-91405 Orsay cedex, France.}

\begin{abstract}
We demonstrate sub-wavelength sectioning on biological samples with a conventional confocal microscope. This optical sectioning is achieved by the phenomenon of supercritical angle fluorescence, wherein only a fluorophore next to the interface of a refractive index discontinuity  can emit propagating components of radiation into the so-called forbidden angles. The simplicity of this technique allows it to be integrated with a high numerical aperture  confocal scanning microscope by only a simple modification on the detection channel. Confocal-SAF microscopy would be a powerful tool to achieve high resolution surface imaging, especially for membrane imaging in biological samples.
 
\end{abstract}

\maketitle %% required

\section*{Introduction}
Confocal microscopy is a mature and widely used tool in biological studies. However, its lateral and axial resolution are limited by diffraction, imposing a limitation on its utility in examining the finer structures present in the cell membrane. The basal membrane of the cell is typically a hotbed of cellular processes involving structure, trafficking, signaling and are widely targeted for drug delivery\cite{hopkins2006}. However, the fine size of the membrane, a few nanometers, precludes the use of confocal microscopy in studying its features exclusively. This is due to the inability of a confocal microscope to discriminate the signal emanating from the membrane as its sectioning capability is in the order of 600 nm. This makes non-invasive optical imaging of cellular membranes particularly challenging.

Among current optical techniques to study the cell membrane, Total Internal Reflection (TIR) microscopy has emerged as a powerful tool that offers superior axial sectioning\cite{axelrod2013}. Due to the evanescent nature of the excitation, only fluorophores in the vicinity of coverslip are selectively imaged, leading to smaller detection volumes. In the commonly used through-the-objective configuration, it suffers from difficulties such as inhomogeneous illumination on the sample and stray excitation. Moreover, in a confocal configuration, there is a significant degradation of the lateral resolution and it is technically challenging to realize diffraction limited focus spots with purely evanescent waves\cite{contirf2004,conftirf2009}. 

An alternate approach to surface-confined imaging is to detect exclusively the supercritical angular fluorescence (SAF) emission\cite{ruckstuhl2000}. When an emitter is sufficiently close to an interface between two media with different refractive indices, the near-field components of its emission can become propagative in the medium with the higher refractive index. This manifests as light emitted into angles well above the critical angle ($\theta_{c} = \arcsin(n_1/n_2))$ where $n_1$ and $n_2$ are the refractive indices on either side of the interface. Such supercritical angular emission can be as high as 34 percent of its total energy for a molecule located on the interface itself and sharply decreases when it is farther away\cite{fort2008}. In imaging biological samples, there exists a sharp refractive index discontinuity at the boundary of the cell, i.e. the membrane itself~($n_{cell}$ = 1.33 - 1.38) and the coverslip-immersion medium ($n_{g} =1.5$), thus fulfilling the conditions necessary for supercritical angular emission.

The discrimination of the supercritical angle fluorescence components (SAF) from the under critical angle fluorescence (UAF) is the key to realize membrane specific imaging. Ruckstuhl et. al have demonstrated a scanning microscope that separated the SAF from the UAF for surface specific imaging\cite{ruckstuhl2004}. This was achieved with a specialized objective with an aspherical lens for excitation and detection of the UAF components and a parabolic collection element that collected exclusively only light emitted at the super-critical angles. While this objective is highly effective in bio-sensing and imaging applications\cite{SAF-biosensor},it offers a relatively lower illumination numerical aperture(N.A. = 1.0)\cite{SAF-nanometric, SAF-simult} than conventional high NA objective lens. Our work is aimed at offering membrane specific imaging on any conventional microscope with relatively minor modification. In this context, we demonstrate the use of a commercially available high N.A. lens to perform surface imaging in a confocal configuration. Contrary to the earlier approach, the discrimination of the SAF and UAF components happens at the back-focal plane of an aplanatic objective lens as in\cite{axelrod2001, barroca2011} for wide-field microscopes. This follows from the Abbe's sine condition which maps the angular emission of a fluorophore to a radial position on the backfocal plane, $ \rho (\theta_{em}) = f_{obj}n_{imm}\sin\theta_{em}$ where $\rho$ is the radial co-ordinate at the back focal plane, $f_{obj}$ is the focal length of the objective, $n_{imm}$ is the refractive index of the immersion medium and $\theta_{em}$ is the emission angle of the dipole. Hence, filtering operations can be conveniently performed on a conjugated plane to separate the SAF and the UAF components of emission. The significant difference to widefield SAF detection with a central mask is that the lateral resolution is not degraded,thereby enabling high resolution surface images with no additional image processing.

\begin{figure}[htbp]
\centerline{\includegraphics[width= \columnwidth]{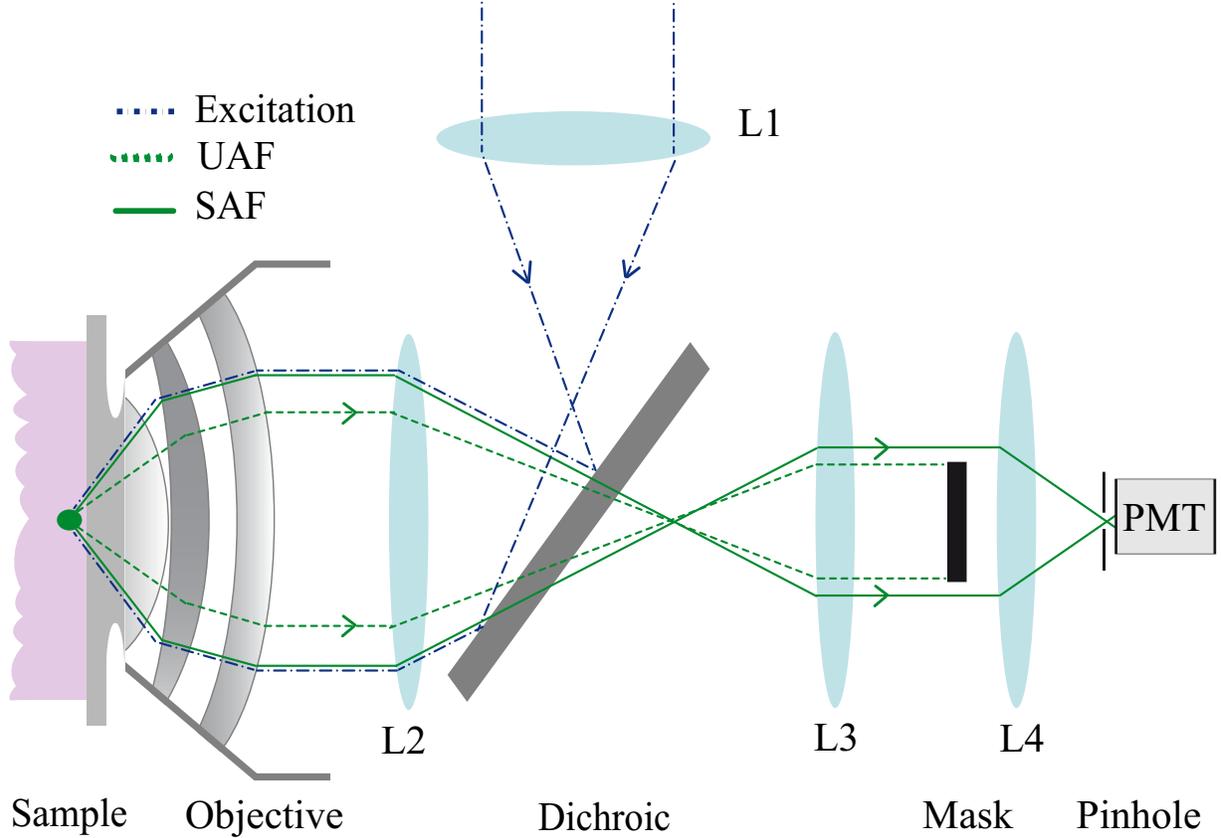}}
\caption{A simplified schematic of a confocal-SAF microscope with an amplitude mask placed in a conjugate plane of the objective's backfocal plane.}
\label{fig:setup}
\end{figure}
A simplified view of the optical setup for SAF imaging is presented in Fig.\ref{fig:setup}. We use an inverted microscope (Ti-E,~\textit{Nikon}) equipped with a 3D piezo-electric stage scanner (P545,~\textit{Physik Instrument} ). The key component of the system is the apochromatic objective lens with a N.A. of 1.49 (VC 1.49 60X,\textit{ Nikon}) offering a collection angle of upto $79.5^{\circ}$ with standard immersion medium ($n_{imm} = 1.515$ ). The output of a pulsed supercontinuum source (SC 450,~\textit{Fianium}) is selected for the excitation wavelength at $488\pm10$~nm, spatially filtered, circularly polarized and expanded by the scan lens (\textit{f}=~80\textit{mm}) and tube lens (\textit{f}=~200\textit{mm}) to overfill the backfocal plane of the objective lens, ensuring a nearly diffraction-limited excitation spot on the sample. Unlike in confocal-TIRF microscopy, this task is relatively simple as the illumination utilizes the entire numerical aperture of the objective lens. The resulting fluorescence emission is then decoupled by a dichroic mirror and relay lenses are used to produce a magnified image ($M \approx 0.48$) of the objective's back-focal plane at an easily accessible location. A central amplitude mask of diameter d $\approx 4.3 mm$ is positioned at this conjugated back-focal plane to block all the light emanating from under the critical angle and ensures only the SAF components propagate further and are focused on to the pinhole with lens L4 (\textit{f}=~300\textit{mm}). A fiber-coupled hybrid-photomultiplier (PMA Hybrid, \textit{Picoquant Gmbh}) running in single photon counting mode is used for the detection and the core of the collection fiber acts as the confocal pinhole. In combination with ideal point like excitation, SAF detection does not require a pinhole for sectioning. However in practice, a large pinhole helps reject any stray light. The detection PSF of SAF components is broader than the conventional Airy disk \cite{sheppardpsf2007}. We measure the focused spot size in SAF detection to be 1.7 times larger than the conventional Airy spot with a CCD camera. We choose a pinhole diameter (105 $\mu$m) that is 1.25 times the projected confocal Airy disk which represents a realistic usage scenario for a confocal  microscope and still collects a large part of the SAF emission. The two detection modes were changed by simply switching the mask in and out of the detection channel. All images were obtained on samples deposited on high precision 0.17 mm coverslips and by recording the SAF and the confocal image sequentially with less than 20 $\mu$W of excitation power at the back-focal plane of the microscope objective.
\begin{figure}[htbp]
\centerline{\includegraphics[width=\columnwidth]{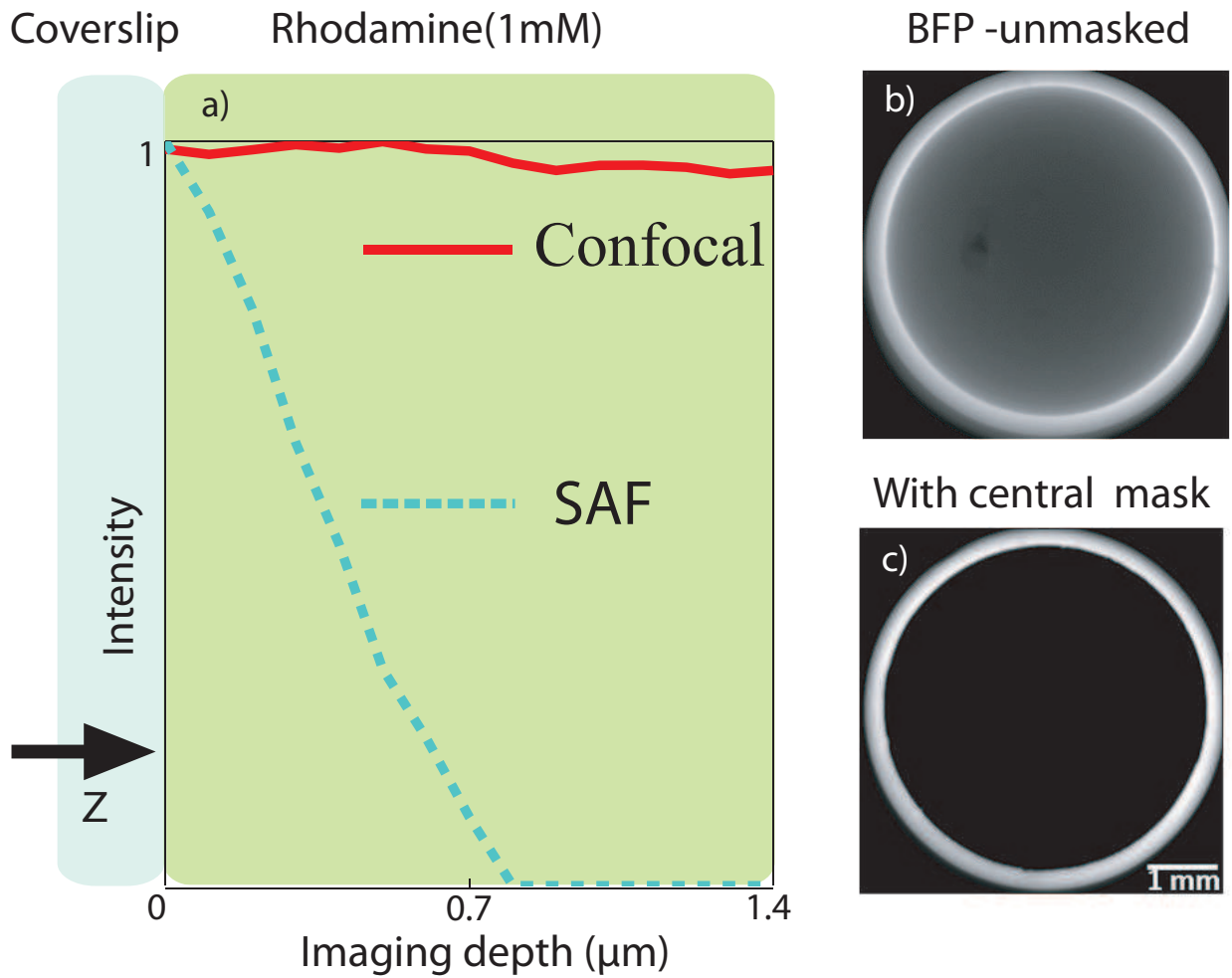}}
\caption{a) Detected fluorescence intensity for an axial scan of the excitation focus in the confocal and SAF modes. The test sample was a 1 mM solution of Rhodamine 6G in ethanol with non-negligible fluorescence in the volume. Intensity counts were obtained for each z step for 30 seconds and normalized with respect to the maximum detected value along the axis. b) An image of the conjugated backfocal plane for a solution of 100 nm beads on the surface, and c) same as b) but with a central mask.}
\label{fig:rhodamine}
\end{figure}

As a test of optical sectioning in the presence of a bright background, we place a 1 mM solution of Rhodamine 6G in ethanol on a coverslip and acquire a set of measured intensity values upto a depth of 1.5 $\mu$m with 100 nm steps. In Fig.\ref{fig:rhodamine}, we clearly see  that while the intensity values in the confocal detection stay constant, the signal in the SAF detection decreases as we probe deeper into the sample despite a large number of fluorophores being excited. We find the axial contrast ratio in this experiment, $C = (I_{max}-I_{min})/(I_{max}+I_{min})$,  along the axis for SAF detection as $C_{SAF}= 0.42$ and for confocal detection as $C_{con}= 0.02$, where $I_{max}$ and $I_{min}$ are the maximum and minimum fluorescence intensity counts recorded in the axial stack for the respective detection mode. The measured intensity curves are not to be misinterpreted as the confinement of the detected volume itself. We briefly describe the formation of an image in SAF detection as
\begin{equation}
Image(r,z) = (PSF(r,z)\otimes Object)\times MCE(z)\times Q
\end{equation}
where $PSF(r,z)$ is the complete 3D point spread function accounting for both excitation and detection conditions, $MCE(z)$ is the molecular collection efficiency for an emitter at a given axial position, $Q$ is the quantum yield and $r, z$ are the lateral and the axial co-ordinates. Image formation in SAF is the product of the light collection efficiency for the supercritical angular components of the emitter at a given depth and its structure convolved with the 3D point spread function of the microscope. The improved sectioning arises not from a narrower axial PSF  but due to the rapid decay of the $MCE$ when the position of the emitter is further away from the interface. The measured curve is only an indication of the axial PSF term (0.8$\mu$m) of the confocal microscope, which is not the dominant factor in sectioning process. However, we clearly see that the detected volume is confined to at least a diffraction limited depth even in the presence of an overwhelming concentration of emitters in the volume.

\begin{figure}[htbp]
\centerline{\includegraphics[width=\columnwidth]{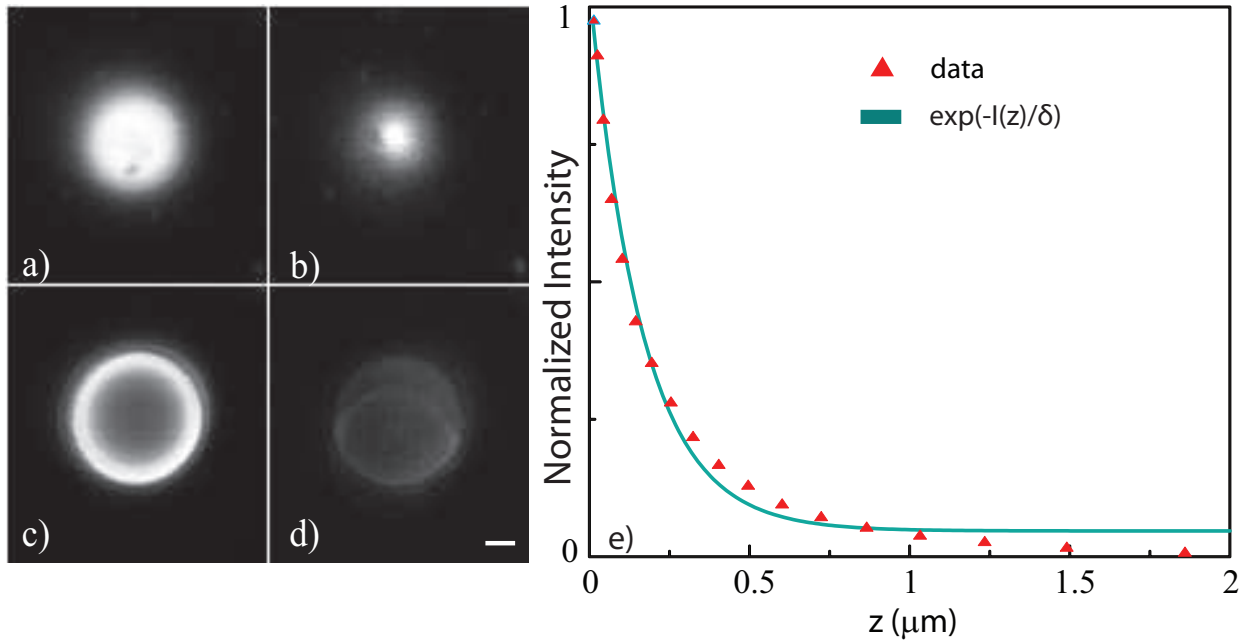}}
\caption{a) and b) are images of the surface planes of a 5 $\mu$m fluorescent bead in confocal and SAF detection modes respectively. c) and d) are the corresponding images 2.5 $\mu$m deep inside the bead imaging its center. The image stack was normalized to the maximum intensity on the surface plane in its corresponding detection mode. e) depicts the intensity decay for emitters located at various depths away from the interface extracted from the radial intensity profile of the surface plane.}
\label{fig:axial}
\end{figure}

\begin{figure}[htbp]
\centerline{\includegraphics[width=\columnwidth]{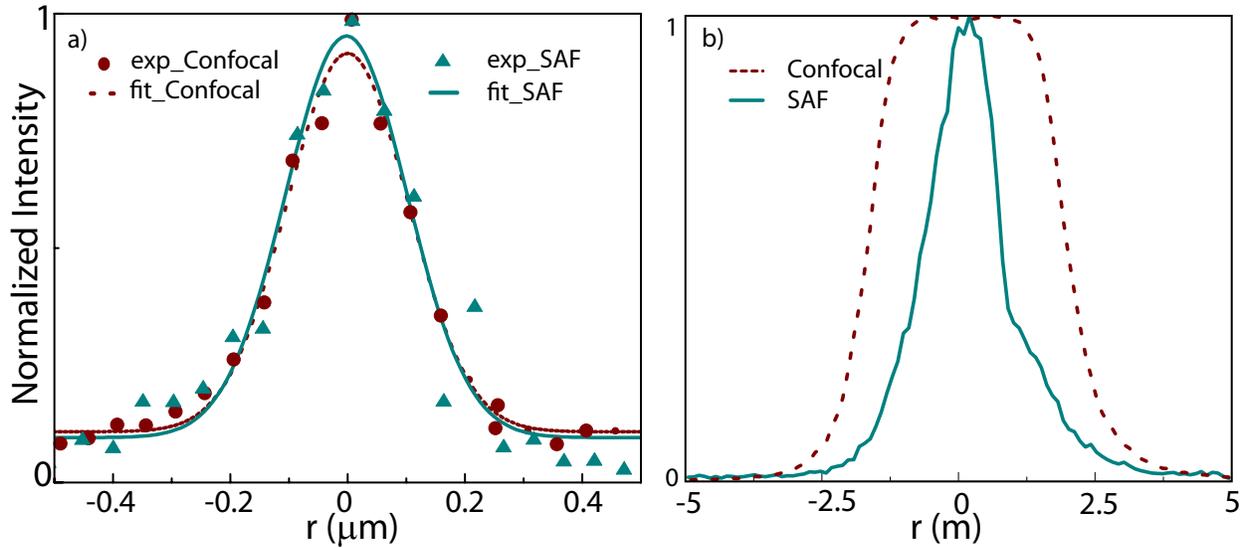}}
\caption{ Normalized lateral profile of a) 100nm beads and b) 5$\mu$m beads on the coverslip in an aqueous solution.}
\label{fig:lateral}
\end{figure}

For quantitative information about the true confinement, the parameter to be estimated is the decay rate of $MCE$ with imaging depth. We acquire a full 3D image stack of 5$\mu$m fluorescent beads (Fig.\ref{fig:axial}). A significant improvement in the images of the surface plane is clearly visible in the SAF detection compared to the confocal detection(Fig.\ref{fig:lateral}). Furthermore, when the focus of the excitation spot is scanned axially, the confocal detection channel records a complete image of the bead while the SAF channel registers only a de-focused image of the surface plane as expected. Following the method introduced by Mattheyses et.al \cite{mattheyses2006}, we analyze the surface image of the bead to deduce the confinement. In a confocal configuration, the pinhole efficiently rejects any stray light that might have been scattered into the super-critical angles. Hence a  mono-exponential decay component is sufficient to model the experimental curve and the confinement is measured to be $167 \pm9 $ nm.

A major drawback of widefield SAF detection is the degradation of the lateral resolution due to the broadening of the detection point spread function and the vSAF technique was introduced to overcome this limitation\cite{vsaf}. However in confocal-SAF, the lateral resolution is determined by the intensity distribution of the excitation light at the focus. Hence, a careful tuning of the objective's correction collar is important to negate the influence of the refractive index mismatch. A solution of 100 nm beads on a coverslip was imaged both in the SAF and confocal modes with a step size of 50 nm and the lateral resolution was determined to be $226\pm 12$~nm in the SAF and $211 \pm 9 $ nm in the confocal mode. The slightly larger PSF can be attributed to the use of the full numerical aperture of the objective (N.A. = 1.49) for excitation in an index mismatched sample, introducing spherical abberation. However its effects are similar in both the modes of imaging and a resolution comparable to conventional confocal microscopes is attained with relatively little experimental effort.

\begin{figure}
\centerline{\includegraphics[width=\columnwidth]{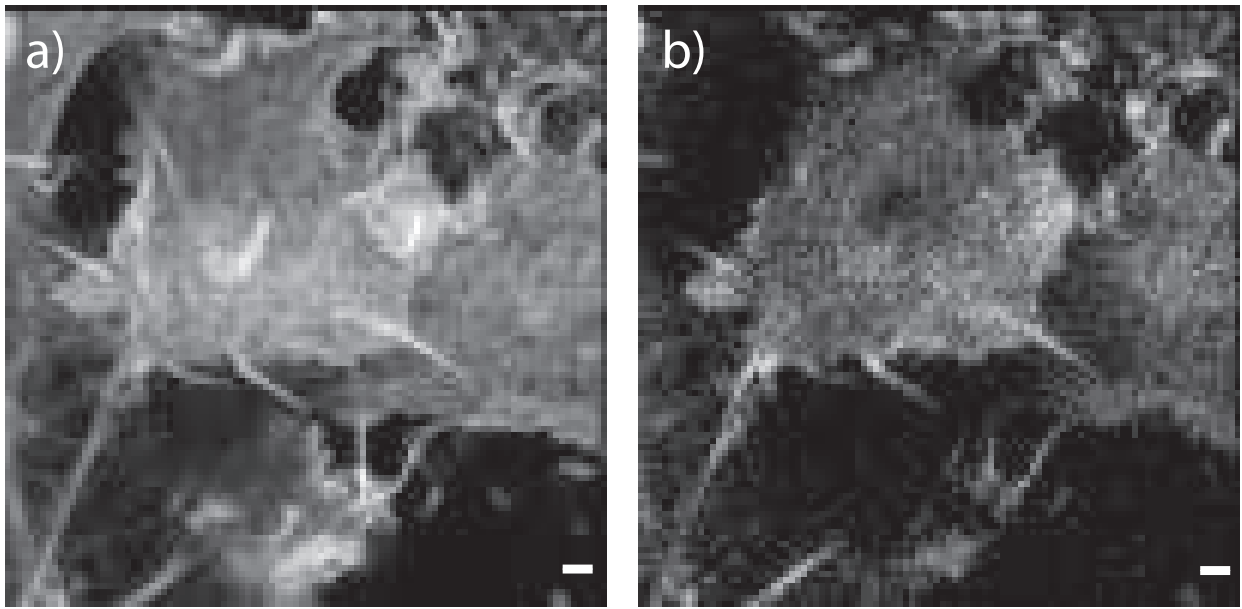}}
\caption{ Actin filaments in fixed CHO cells tagged with ATTO 488 dye in a) confocal detection and b) confocal-SAF detection. (Scale bar -1$\mu m$, Field of View:20$\mu m \times 20 \mu m$)}
\label{fig:cells}
\end{figure}

In order to demonstrate the application of confocal-SAF microscopy in biological samples, we image actin filaments in fixed CHO cells stained with ATTO488 dye. Actin microfilaments are largely present in the cytoskeleton of the cell and closely interact with the cellular membranes governing several processes involving motility and signalling. We clearly see its features preserved in the SAF image while light from the volume of the cell is rejected. Also, to ensure these vanishing features in the SAF image are not artifacts, we acquired a 3D stack and verified in the confocal images that these regions corresponded to actin filaments extending deeper into the cell. The speed of acquisition was limited to 2 ms per pixel due to the piezo-electric scanner. However, introduction of well-positioned scanning mirrors could easily facilitate faster image acquisition. In future, we intend to extend our work to realize simultaneous detection of both the UAF and SAF components to realize tomographic sectioning based on the ratio of detected intensities.

In this letter, we have demonstrated the potential of confocal-SAF microscopes to detect highly surface specific features with high lateral resolution. An important implication of using conventional microscope optics is the ability to combine SAF detection in applications requiring more freedom on the illumination side, like PSF engineering, wavefront and polarization coding, tighter focal spots like multi-photon excitation and super-resolution techniques, fluorescence correlation spectroscopy and multi-colour imaging. This would enable a wide array of multi-modal imaging configurations in the future. The remarkable simplicity and ease of the technique makes it a powerful optical tool for non-invasive surface imaging.

The authors thank S.Lecart for helpful comments on the article. This work has been supported by grants from the L'Agence Nationale de la Recherche (ANR STEDFLIM, ANR SMARTVIEW) and RTRA Triangle de la Physique. E.F.thanks AXA Research Fund for financial support.

\newpage
\section*{Informational fifth page}

\end{document}